\documentclass{iopconfser}
\usepackage{graphicx}
\usepackage{cite}

\begin{document}

\title{Implicit PIC Simulations of Hydrogen Plasmas with Intense Negative Ion Production}

\author{D. Eremin$^{1}$}

\affil{$^1$Institute of Theoretical Electrical Engineering, Ruhr University Bochum, Bochum, Germany}

\email{denis.eremin@rub.de}

\begin{abstract}
High-density plasma discharges in hydrogen operating at low pressures are of considerable interest for negative ion production used in many applications. Their weak collisionality requires a kinetic, nonlocal description, and a popular approach is to model such discharges numerically using the particle-in-cell (PIC) method combined with Monte Carlo collisions (MCC). However, the most commonly used explicit momentum-conserving variant of this method is highly inefficient because it must resolve both the Debye length and the plasma period. In this context, the implicit energy-conserving PIC/MCC scheme appears to be an attractive alternative, as it can model the plasmas of interest using their actual parameters without resorting to artificial scaling techniques. This work considers simulations of two types of high-density plasma discharges performed using electromagnetic or electrostatic energy-conserving PIC/MCC codes in hydrogen, and discusses the observed negative ion production.
\end{abstract}

\section{Introduction}

Hydrogen plasmas are of interest to many applications related to plasma processing, lithography machines, or negative ion sources utilized in thermonuclear fusion or particle accelerators. Many of the corresponding discharges feature high density plasmas operating at a low pressure, which requires taking into account kinetic and nonlocal effects when modeling them. A very popular approach to meet this goal is to use the particle-in-cell method enhanced by Monte Carlo algorithm for the description of collisions \cite{Birdsall1991} that models plasmas virtually from the first principles, describing discharges in question as a self-consistently evolving system of colliding charged particles and electromagnetic fields. However, to cut down computation costs, large clusters of particles with a similar location in phase space are ``merged'' into superparticles, whose number is much smaller compared to the number of original particles, whereas the fields are discretized on a fixed Eulerian grid, and the field equations are influenced by the superparticles through charge and/or currrent densities. To calculate the latter on the grid and to determine electromagnetic forces acting on particles at their exact locations, an interpolation procedure from/to gridded quantities to continuous coordinates must be defined. Such an interpolation involves a ``shape function'', and it can be proven that if the shape function is the same for the electric field and the charge density, the corresponding PIC algorithm conserves momentum, while choosing the same shape function for the electric field and current density ensures energy conservation. It was recently discovered that the momentum and energy conserving PIC algorithms have rather different criteria for the onset of so-called finite-grid-instability (FGI). FGI emerges due to the fact that superparticles move under the influence of electric field, which has discontinuities of its value or value of its derivatives at the grid nodes. This entails spatial harmonics in the charge or current densities, which may couple to the original spatial field harmonics, resulting in FGI under certain criteria. For the momentum-conserving PIC FGI arises if cell size is greater than a quantity of the order of the Debye length, and the energy-conserving PIC is stable with respect to FGI provided that the average electron drift velocity is smaller than its thermal velocity \cite{Barnes2021}. Whereas in the first case the stability criterion is strongly restrictive for high density plasmas, in the second case PIC simulations remain FGI-stable for most of plasmas. This is why the energy-conserving PIC algorithm augmented with nonuniform grids enables simulations of high density plasma discharges with the original parameters, whereas simulations of such discharges with the momentum-conserving PIC is tractable only if additional tricks are used, such as, for example, scaling techniques \cite{li_2019}. Note, however, that such techniques may not work well if investigated plasmas exhibit nonlinear processes, such as binary collisions involving species in PIC or various instabilities, which is often the case.

Another important property of a PIC algorithm is its approach to integration of the particle motion and field evolution equations in time. In the vast majority of cases such an integration is based on the leapfrog scheme, which relies on utilizing information available from the previous time levels, making the time integration algorithm explicit. In this case, evolution of the charged particles and electromagnetic fields are decoupled from each other during the current time step, and any unresolved time scales related to oscillations or waves inherent to plasmas can make the algorithm unstable. 
For example, electromagnetic waves in plasma, which have dispersion relation $\omega^2 = \omega_{pe}^2 + k^2c^2$, lead to the demand to resolve the plasma frequency in the long-wavelength (or electrostatic) limit $k\to 0$ or the Courant-Friedichs-Lewy stability condition $\Delta t < \Delta x/c$ in the short-wavelength limit $k \to 2\pi/\Delta x$. In contrast, time integration based on Crank-Nicolson approach, which is implicit, entangling the old and new quantities and making evolution of particle orbits and electromagnetic fields self-consistent during the current time step, is unconditionally stable. 
However, in practice, the time step size is typically determined by the convergence of the iterative procedure used to solve the implicit equations, the accuracy required for particle orbit integration, and the characteristic time scales of the plasma dynamics. Excessively large time steps may necessitate subcycling of the particle orbit integration, thereby increasing the computational cost, or may severely degrade the convergence of the iterative procedure unless suitable preconditioners are employed \cite{chen_2014}.

The present work applies the ECCOPIC (Energy- and Charge-Conserving PIC) family of codes constructed in-house, verified and validated in \cite{charoy_2019,Eremin2023,Berger2023,Eremin2025}, to model two distinct plasma discharges - a surfatron and a capacitively-coupled plasma (CCP) discharge in hydrogen at 5 Pa with a novel feature, which takes into account self-consistent dynamics of negative hydrogen ions. The paper is organized as follows: Section \ref{s2} describes numerical models used, Results of numerical simulations are discussed in Section \ref{s3}, and Section \ref{s4} draws conclusions.

\section{Models \label{s2}}

Since electromagnetic effects are important in surfatron discharges, the fields are calculated by solving the fully electromagnetic Maxwell equations. In contrast, the CCP discharge can be adequately described using the electrostatic approximation. In both cases, charge conservation is ensured by accounting for cell-boundary crossings of superparticles. Consequently, enforcing Gauss's law through the Poisson equation is unnecessary in the electromagnetic case, whereas in the electrostatic case, the Poisson equation can be used to calculate the electrostatic potential instead of solving the divergence of Ampère's law.

\subsection{Surfatron Discharge}

\begin{figure}[h]
\centering
\includegraphics[width=5cm]{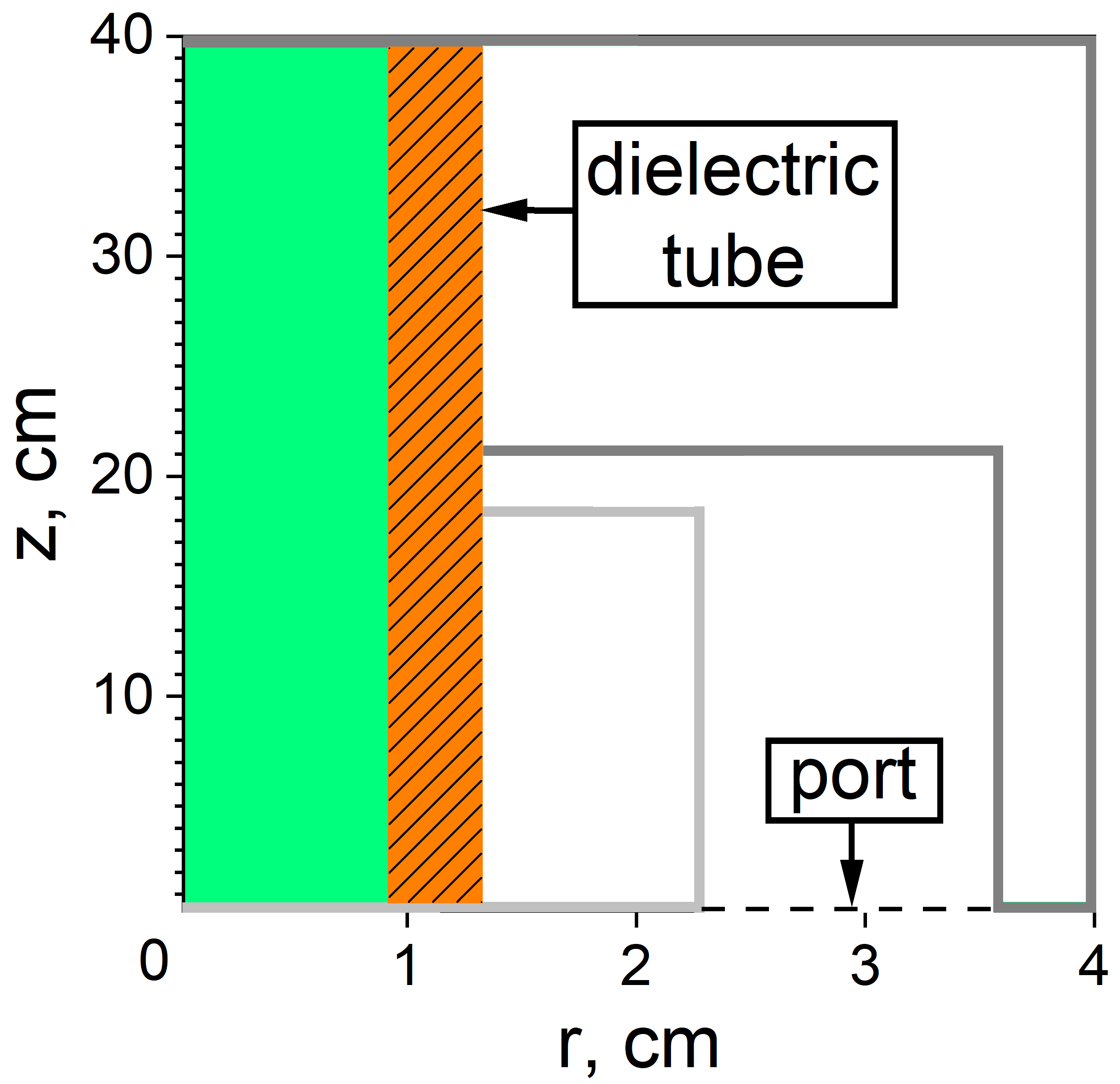}
\caption{Model geometry of the surfatron discharge. The discharge, ignited in the reactor tube shown in green, is powered by the TEM$_{00}$ wave entering through the port between two metal cylinders (light gray and dark gray) at the bottom, which is converted into a surface wave at the quartz dielectric tube (orange). The cylindrical axis is pointing in the vertical direction.}
\label{Fig1}
\end{figure}

The electromagnetic energy-conserving PIC schemes for simulations of the surfatron discharge is based on the following equations. Assuming cylindrical geometry with the azimuthal symmetry, the appropriate field pattern includes the azimuthal magnetic field along with the radial and axial electric field components \cite{lieberman_2002,sansonnens_2006,Eremin2025,Eremin2025a}

\begin{equation}
\left\{
\begin{array}{lll}
\mu \frac{\partial}{\partial t} H^\theta &=& \frac{1}{J}\left(\frac{\partial}{\partial\xi} E_\eta - \frac{\partial}{\partial \eta} E_\xi\right) \\
\left(\sigma+\epsilon \frac{\partial}{\partial t}\right) E^\xi &=& -j^\xi - \frac{1}{J}\frac{\partial}{\partial\eta} H_\theta \\
\left(\sigma+\epsilon \frac{\partial}{\partial t}\right) E^\eta &=& -j^\eta + \frac{1}{J}\frac{\partial}{\partial \xi} H_\theta,
\end{array}
\right. \label{eq2_1}
\end{equation}
and the field components are discretized according to the Yee scheme \cite{yee_1966}. In this equation, quantities with the superscript (subscript) denote the contravariant (covariant) field components \cite{chacon_2013}, $J_r = dr/d\xi$ with $\xi$ the logical radial coordinate, $J_z = dz/d\eta$ with $\eta$ the logical axial coordinate, the field grid in both logical coordinates being equidistant, and $J = J_rJ_z$. Note that $E_{\xi} = J_r^2 E^{\xi}$, $E_{\eta} = J_z^2 E^{\eta}$, and $H_\theta = r^2 H^\theta$.  

To complete the field equations, one needs to specify a formula for the current density, which in the considered case of a two-dimensional problem is provided by \cite{chacon_2013}
\begin{equation}
\left\{
\begin{array}{lll}
j^{\xi}_{s;i+1/2,j} &=& \frac{q_s}{J\Delta\xi\Delta\eta}\sum\limits_p w_p  v_{rp} J_r(\xi_p) S_{m-1}(\xi_p-\xi_{i+1/2}) S_m(\eta_p-\eta_j) \\ 
j^{\eta}_{s;i,j+1/2} &=& \frac{q_s}{J\Delta\xi\Delta\eta}\sum\limits_p w_p v_{zp} J_z(\eta_p) S_{m}(\xi_p-\xi_i) S_{m-1}(\eta_p-\eta_{j+1/2}),
\end{array} \label{eq2_2}
\right.
\end{equation}
where $\Delta \xi$ and $\Delta \eta$ denote the cell sizes of the logical grid, $S_m$ is the B-spline of order $m$ \cite{chen_2011}, $p$ subscript denotes contribution from the $p$th particle, $w_p$ is the superparticle weight (number of real particles in $p$th superparticle), $q_s$ is the charge of particle species $s$, and $i$ or $j$ denote the vertices of the corresponding field grid. The charge conservation is ensured by calculating the current densities each time a particle crossed a cell boundary \cite{chen_2011}. When the same interpolation functions are used to evaluate the electric field at particle positions from its grid-based representation, energy conservation can be ensured to a prescribed numerical precision in a bounded, collisional plasma, thereby satisfying the exact energy balance equation \cite{Eremin2022}.

For the given situation, equations of particle motion in cylindrical coordinates read (e.g., \cite{delzanno_2013b}),
\begin{equation}
\left\{
\begin{array}{rll}
m_s\frac{dv_r}{dt} &=& \frac{q_sE_{\xi}}{J_r} + \frac{m_sv_\theta^2}{r} \\
m_s\frac{dv_\theta}{dt} &=& -\frac{m_sv_r v_\theta}{r} \\
m_s\frac{dv_z}{dt} &=& \frac{q_sE_\eta}{J_z}\\
\frac{d\xi}{dt} &=& \frac{v_r}{J_r}\\
\frac{d\eta}{dt} &=& \frac{v_z}{J_z} , 
\end{array}
\right.   \label{eq2_3}
\end{equation}
where particle velocities are $v_r = dr/dt$ and $v_z=dz/dt$ and the positions
are evolved in the logical coordinates \cite{chacon_2013,wang_1999,delzanno_2013}. Here, we assumed that the Lorentz terms caused by the magnetic field of the wave is negligibly small. 

Eqs. (\ref{eq2_1},\ref{eq2_2},\ref{eq2_3}) must be discretized in time. To this end, the time derivatives should be replaced by $\partial_t f = (f^{n+1}-f^n)/\Delta t$, and all other quantities should be taken at the mid-level between the new and the old time levels, $f^{n+1/2} = (f^n+f^{n+1})/2$. In calculation of the current density components in Eq.(\ref{eq2_2}) and analogous expressions for interpolation of the electric field components to the particle location we used $S^{n+1/2}_{m-1}(f_p-f_g) \equiv S_{m-1}((f^{n+1}_p+f^n_p)/2-f_g)$, and $S^{n+1/2}_{m}(f_p-f_g) \equiv (S_m(f^{n+1}_p-f_g)+S_m(f^n_p-f_g))/2$ with $f$ denoting the appropriate coordinate, $p$ particle index, and $g$ 
grid location. 

The boundary condition at the entrance port (filled with dielectric in the ECR discharge and vacuum or air in the surfatron discharge) is formulated in the form of the Mur condition for the outgoing TEM wave, $(J_z^{-1}\partial/\partial \eta - (\sqrt{\epsilon\mu}/c) \partial/\partial t)H_\theta^- = 0$ \cite{mur_1981}. Then, recalling that $H_\theta^-=H_\theta - H_\theta^+$ with $H_\theta$ the total magnetic field intensity and $H_\theta^+$ the incoming wave at $\eta=0$ described by $H_\theta^+ = A(t)\cos(\omega t)$ with $A(t)$ slowly varying amplitude, one
can write the boundary condition for $H_\theta$ at the port as \cite{alvarez_2007,jimenez-diaz_2011,rahimi_2014}
\begin{equation}
\left(\frac{1}{J_z}\frac{\partial}{\partial \eta}  - \frac{\sqrt{\epsilon\mu}}{c}\frac{\partial}{\partial t} \right) H_\theta = Q \label{eq2_4}
\end{equation}
with the source $Q$ specified by the prescribed incoming wave, $Q = -2 (\sqrt{\epsilon\mu}/c)\partial H^+_\theta/\partial t$. In this way the outgoing wave is allowed to leave the computational domain virtually without reflections at the port so that the radiation, which was not absorbed by plasma, is not accumulated in the discharge. The amplitude $A(t)$ is adjusted to match the period-averaged power absorbed in plasma \cite{eremin_2016b} via $A^{N+1} = A^N (1-\alpha(P_N-P_0)/P_0$, where $P_0$ is the desired power, the absorbed power is calculated as $P_N = 2\pi \int^{N T_{MW}}_0 dt \int^R_0 rdr\int_0^Z dz (j_rE_r+j_zE_z)/(NT_{MW})$ with $N$ number of MW periods $T_{MW}$ for the period averaging, and $\alpha$ is a small numerical parameter. For simulations conducted in the present work $N$ and $\alpha$ were chosen to be $100$ and $0.1$, respectively.

\subsection{CCP Discharge}

\begin{figure}[h]
\centering
\includegraphics[width=5cm]{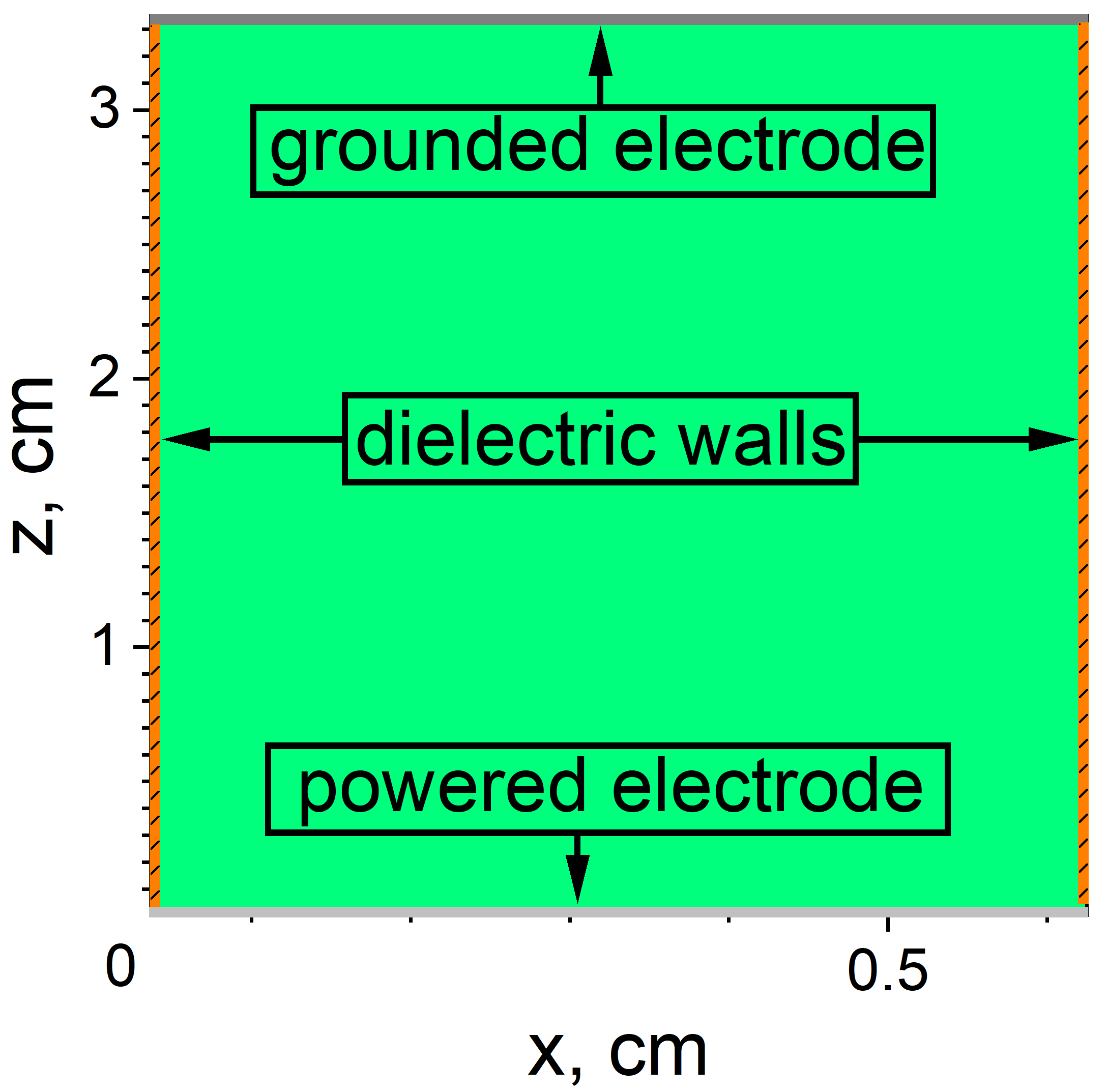}
\caption{Model geometry of the CCP discharge. The discharge, ignited in the semi-infinite reactor chamber with rectangular cross-section shown in green, is powered by the electric field created between two metal electrodes, powered one at the bottom with a voltage applied to it (light gray), and a grounded one at the top (dark gray). On the sides the discharge is bounded by dielectric walls shown in orange.}
\label{Fig2}
\end{figure}

Equations for the electrostatic version of the energy-conserving PIC that was used to model the CCP reactor in this work can be obtained from Eqs.(\ref{eq2_2}) and (\ref{eq2_3}) in the following way. First of all, due to the rectangular geometry the radial $r$ and axial $z$ coordinates are replaced by Cartesian coordinates $x$ and $z$. 
The azimuthal coordinate is replaced by Cartesian coordinate $y$, and we assume translational invariance in this direction. Due to the rectangular geometry the last term in each of the top two lines of Eqs.(\ref{eq2_3}) is absent. The electric field at the time level $n+1$ in the energy-conserving PIC method is supposed to be calculated from the divergence of Ampère's law. The latter can be deduced from the bottom two lines in Eq.(\ref{eq2_1})) adapted to the chosen geometry. If in addition charge conservation is ensured, one can equivalently use the Poisson equation \cite{chen_2011,eremin_2021}. In the logical coordinates $\xi = \xi(x)$ and $\eta = \eta(z)$ the electrostatic potential can therefore be determined from the following equation:
\begin{equation}
\frac{1}{J_x}\frac{\partial}{\partial \xi} \frac{1}{J_x} \frac{\partial}{\partial \xi} \phi + \frac{1}{J_z}\frac{\partial}{\partial \eta} \frac{1}{J_z} \frac{\partial}{\partial \eta} \phi = -\sum\limits_s \rho_s \label{eq2_5}
\end{equation}
with $J_x = dx/d\xi$, $J_z=dz/d\eta$, and
\begin{equation}
\rho_{s;i,j} = \frac{q_s}{J_xJ_z\Delta\xi\Delta\eta}\sum\limits_p w_p  S_{m}(\xi_p-\xi_i) S_m(\eta_p-\eta_j). 
\label{eq2_6}
\end{equation}
The boundary conditions adopted in this case set $V(t)=V_0\sin(\omega t)$ with a fixed amplitude $V_0$ at the bottom electrode, $0$ at the top electrode, and $E_\xi = 0$ at the left and right sides.


\subsection{Hydrogen Chemistry Model}

\begin{table}[t!]
        \footnotesize
		\centering
		\begin{tabular*}{1.0\textwidth}{ l @{\extracolsep{\fill}} l l c c}
			\hline
			No.~ & reaction & process name & $\varepsilon_\mathrm{thr}, \mathrm{eV}$ &  src. \\
			\hline
			R1 & $\mathrm{e} + \mathrm{H_2} \rightarrow \mathrm{e} + \mathrm{H_2}$ & elastic scattering & - & \cite{brunger_2002} \\
			R2 & $\mathrm{e} + \mathrm{H_2} \rightarrow 2\mathrm{e} + \mathrm{H_2^+}$ & electron impact ionization & $15.43$ & \cite{janev2003} \\
			R3 & $\mathrm{e} + \mathrm{H_2} \rightarrow 2\mathrm{e} + \mathrm{H_2^+}\,(X\, ^2\Sigma_{\mathrm{g}^+}) \rightarrow 2\mathrm{e} + \mathrm{H^+} + \mathrm{H}$ & dissociative ionization & $18.46$ & \cite{janev2003} \\
			R4 & $\mathrm{e} + \mathrm{H_2} \rightarrow 2\mathrm{e} + \mathrm{H_2^+}\,(B\, ^2\Sigma_{\mathrm{u}^+}) \rightarrow 2\mathrm{e} + \mathrm{H^+} + \mathrm{H}$ & dissociative ionization & $32.69$ & \cite{janev2003} \\
			R5 & $\mathrm{e} + \mathrm{H_2}(j=0) \rightarrow \mathrm{e} + \mathrm{H_2}(j=2)$ & rotational excitation & 0.044 & \cite{england_1988} \\
		      R6 & $\mathrm{e} + \mathrm{H_2}(j=1) \rightarrow \mathrm{e} + \mathrm{H_2}(j=3)$ & rotational excitation & 0.073 & \cite{england_1988} \\
			R7 & $\mathrm{e} + \mathrm{H_2} \rightarrow \mathrm{e} + \mathrm{H_2}\,(B\, ^1\Sigma_{\mathrm{u}}^+)$ & electronic excitation & 11.3 & \cite{lxcat} \\
			R8 & $\mathrm{e} + \mathrm{H_2} \rightarrow \mathrm{e} + \mathrm{H_2}\,(C\, ^1\Pi_{\mathrm{u}})$ & electronic excitation & 12.4 & \cite{lxcat} \\
			R9 & $\mathrm{e} + \mathrm{H_2} \rightarrow \mathrm{e} + \mathrm{H_2}\,(a\, ^3\Sigma_{\mathrm{g}}^+)$ & electronic excitation & 11.8 & \cite{lxcat} \\
			R10 & $\mathrm{e} + \mathrm{H_2} \rightarrow \mathrm{e} + \mathrm{H_2}\,(c\, ^3\Pi_{\mathrm{u}})$ & electronic excitation & 11.72 & \cite{lxcat} \\
			R11 & $\mathrm{e} + \mathrm{H_2} \rightarrow \mathrm{e} + \mathrm{H_2}\,(d\, ^3\Pi_{\mathrm{u}})$ & electronic excitation & 14.0 & \cite{lxcat} \\
			R12 & $\mathrm{e} + \mathrm{H_2} \rightarrow \mathrm{e} + \mathrm{H_2}\,(e\, ^3\Sigma_{\mathrm{u}}^+)$ & electronic excitation & 13.0 & \cite{yoon_2008} \\
			R13 & $\mathrm{e} + \mathrm{H_2} \rightarrow \mathrm{e} + \mathrm{H_2}\,({\rm Rydberg, sum})$ & electronic excitation & 16.0 & \cite{lxcat} \\ 
			R14 \,\, & $\mathrm{e} + \mathrm{H_2} \rightarrow \mathrm{e} + 2\mathrm{H} $ & dissociation & 7.93 & \cite{yoon_2008} \\   
			\hline
		\end{tabular*}
		\caption{List of reactions between electrons and molecular hydrogen. The second column from the right indicates threshold energies $\varepsilon_\mathrm{thr}$.}
		\label{tab:chemistry1}
\end{table}

\begin{table}[t!]
        \footnotesize
		\centering
		\begin{tabular*}{1.0\textwidth}{ l @{\extracolsep{\fill}} l l c c}
			\hline
			No.~ & reaction & process name & $\varepsilon_\mathrm{thr}, \mathrm{eV}$ &  src. \\
			\hline
			R15 & $\mathrm{H^+} + \mathrm{H_2} \rightarrow \mathrm{H^+} + \mathrm{H_2}$ & elastic scattering & - & \cite{simko_1997} \\
			R16 & $\mathrm{H^+} + \mathrm{H_2} (j=0) \rightarrow \mathrm{H^+} + \mathrm{H_2}(j=2)$ & rotational excitation & 0.044 & \cite{simko_1997} \\
			R17 & $\mathrm{H^+} + \mathrm{H_2} (j=1) \rightarrow \mathrm{H^+} + \mathrm{H_2}(j=3)$ & rotational excitation & 0.077 & \cite{simko_1997} \\
			R18 & $\mathrm{H^+} + \mathrm{H_2}(v=0) \rightarrow \mathrm{H^+} + \mathrm{H_2}(v=1)$ & vibrational excitation & 0.516 & \cite{simko_1997} \\
			R19 & $\mathrm{H^+} + \mathrm{H_2}(v=0) \rightarrow \mathrm{H^+} + \mathrm{H_2}(v=2)$ & vibrational excitation & 1.0 & \cite{simko_1997} \\
			R20 & $\mathrm{H^+} + \mathrm{H_2}(v=0) \rightarrow \mathrm{H^+} + \mathrm{H_2}(v=3)$ & vibrational excitation & 1.5 & \cite{simko_1997} \\  
			R21 & $\mathrm{H^+} + \mathrm{H_2} \rightarrow \mathrm{H} + \mathrm{H_2^+}$ & charge transfer & 2.03 & \cite{simko_1997} \\   
			R22 & $\mathrm{H_2^+} + \mathrm{H_2} \rightarrow \mathrm{H_3^+} + \mathrm{H}$ & formation of $\mathrm{H_3^+}$ & -1.71 & \cite{phelps_1990} \\
			R23 & $\mathrm{H_2^+} + \mathrm{H_2} \rightarrow \mathrm{H_2} + \mathrm{H_2^+}$ & charge transfer & - & \cite{phelps_1990} \\
			R24 & $\mathrm{H_2^+} + \mathrm{H_2}(v=0) \rightarrow \mathrm{H_2^+} + \mathrm{H_2}(v=1)$ & vibrational excitation & 0.516 & \cite{phelps_1990} \\
			R25 & $\mathrm{H_2^+} + \mathrm{H_2} \rightarrow \mathrm{H^+} + \mathrm{H} + \mathrm{H_2}$ & dissociation & 2.65 & \cite{phelps_1990} \\   
			R26 & $\mathrm{H_3^+} + \mathrm{H_2} \rightarrow \mathrm{H_3^+} + \mathrm{H_2}$ & elastic scattering & - & \cite{simko_1997} \\ 
			R27 & $\mathrm{H_3^+} + \mathrm{H_2} \rightarrow \mathrm{H_2} + \mathrm{H} + \mathrm{H^+} + \mathrm{H}$ & dissociative charge transfer & 8.84 & \cite{peko_1997} \\ 
			R28 & $\mathrm{H_3^+} + \mathrm{H_2} \rightarrow \mathrm{H_2} + \mathrm{H} + \mathrm{H_2^+}$ & dissociative charge transfer & 6.19 & \cite{peko_1997} \\ 
			R29 & $\mathrm{H_3^+} + \mathrm{H_2} \rightarrow \mathrm{H^+} + \mathrm{H_2} + \mathrm{H_2}$ & dissociation & 4.36 & \cite{peko_1997} \\ 
			R30 & $\mathrm{H_3^+} + \mathrm{H_2} \rightarrow \mathrm{H_2^+} + \mathrm{H} + \mathrm{H_2}$ & dissociation & 6.19 & \cite{phelps_1990} \\ 
			R31 & $\mathrm{H} + \mathrm{H_2} \rightarrow \mathrm{H} + \mathrm{H_2}$ & elastic scattering & - & \cite{simko_1997} \\ 
			\hline
		\end{tabular*}
		\caption{List of reactions between heavy particles and molecular hydrogen. The second column from the right indicates threshold energies $\varepsilon_\mathrm{thr}$.}
		\label{tab:chemistry2}
\end{table}

\begin{table}[t!]
        \footnotesize
		\centering
		\begin{tabular*}{1.0\textwidth}{ l @{\extracolsep{\fill}} l l c c}
			\hline
			No.~ & reaction & process name & $\varepsilon_\mathrm{thr}, \mathrm{eV}$ &  src. \\
			\hline
			R32 & $\mathrm{e} + \mathrm{H_2}(v=0) \rightarrow \mathrm{e} + \mathrm{H_2}(v=1,2,3)$ & eV vibrational excitation & 0.516 (v=1), 1.0 (v=2), 1.5 (v=3) & \cite{ehrhardt_1968} \\
		      R33 & $\mathrm{e} + \mathrm{H_2}(v=0) \rightarrow \mathrm{e} + \mathrm{H_2}(v=0-14)$ & EV vibrational excitation & $\approx$ 20 & \cite{hiskes_1991} \\
		      R34 & $\mathrm{e} + \mathrm{H_2}(v=0-11) \rightarrow \mathrm{H^-} + \mathrm{H}$ & dissociative attachment & 3.73 (v=0) ... 0 (v=11)  & \cite{janev2003} \\ 
            R35 & $\mathrm{e} + \mathrm{H^-} \rightarrow 2\mathrm{e} + \mathrm{H}$ & electron detachment & 1.55 & \cite{janev2003} \\  
            R36 & $\mathrm{H} + \mathrm{H^-} \rightarrow \mathrm{e} + \mathrm{H_2}$ & associative detachment & - & \cite{janev2003} \\        
			\hline
		\end{tabular*}
		\caption{List of reactions between electrons and molecular hydrogen, which are related to the negative ion chemistry. The second column from the right indicates threshold energies $\varepsilon_\mathrm{thr}$.}
		\label{tab:chemistry3}
\end{table}

The major part of the chemistry set follows the one used by the author in a previous work \cite{Eremin2025}, see Tables 1 and 2. However, since it did not include negative ions, this work extends the previous set to account for those, see Table 3. 

The dominant mechanism of H$^{-}$ creation is the dissociative attachment reaction R34, which occurs when electrons collide with vibrationally excited hydrogen molecules. The vibrational state populations were calculated using the reduced linear model (RLM) developed in \cite{yang_2019}. The RLM assumes that the vibrational state populations are determined by the balance between electron-impact excitation from the ground state to vibrationally excited states, occurring either directly (the eV process, R32; only the first three vibrational levels are included because the corresponding cross sections decrease rapidly with increasing vibrational level) or indirectly via higher singlet states (the EV process, R33), and vibrational de-excitation to lower states through collisions with the discharge walls. Since the mean free path of molecular hydrogen at room temperature and pressure of 5 Pa is approximately 3 mm, which is of the same order of magnitude as the reactor tube radius in surfatron and the reactor width between dielectric walls in the CCP discharge (9 and 6.25 mm, respectively), this ansatz appears to be reasonable. The key assumption of the RLM is that the repopulation probability of a vibrational level $v$ from a higher vibrational level $v^\prime$ due to a wall collision can be approximated as $\gamma_{v^\prime\to v} = (1/v^\prime)$. When no loss of the excited molecules due to a possible outflow from the reactor chamber is considered, the density of vibrational level $v$ can be calculated from the following formula \cite{yang_2019}:
\begin{equation}
n_{H_2(v)} = \left( S_v + \frac{1}{v+1}\sum\limits_{k=v+1}^N S_k \right)/k_{{\rm WR},v},
\end{equation}
where $S_v$ is the rate of electron-impact excitations of a vibrational level $v$ due to the eV and EV processes, which is evaluated in the PIC simulations directly, and $k_{{\rm WR},v}$ is the wall repopulation rate of the vibrational level $v$, which is given by  
\begin{equation}
k_{{\rm WR},v} = \left[ \frac{\Lambda^2}{D} + \frac{2V}{A{\rm v}_{th}}\left( \frac{1+R_v}{1-R_v}\right) \right]^{-1},
\end{equation}
with $\Lambda$ the diffusion scale of the reactor chmaber of volume $V$ and surface area $A$ (due to the high anisotropy of the reactor geometry in both the surfatron and CCP reactors the radial and horizontal sizes were assumed as $\Lambda$ for the corresponding cases), $D$ the diffusion coefficient, ${\rm v}_{th}$ the thermal velocity, and $R_v$ the reflection coefficients provided by Hiskes and Karo in \cite{hiskes_1989}. 

In the present model H$^{-}$ ions are assumed to be destroyed primarily via electron and associative detachment reactions, R35 and R36, respectively. As the electron detachment appears to be a stronger source of negative ion losses, if one succeeds in weakening this reaction through lowering the electron temperature (the corresponding cross-section has threshold at 1.55 eV and grows until it reaches peak at 17.8 eV), it should lead to increase of the negative ion density. This will be important for a later discussion.

\vskip 0.5cm
\section{Results \label{s3}}

\begin{figure}[h]
\centering
\includegraphics[width=12cm]{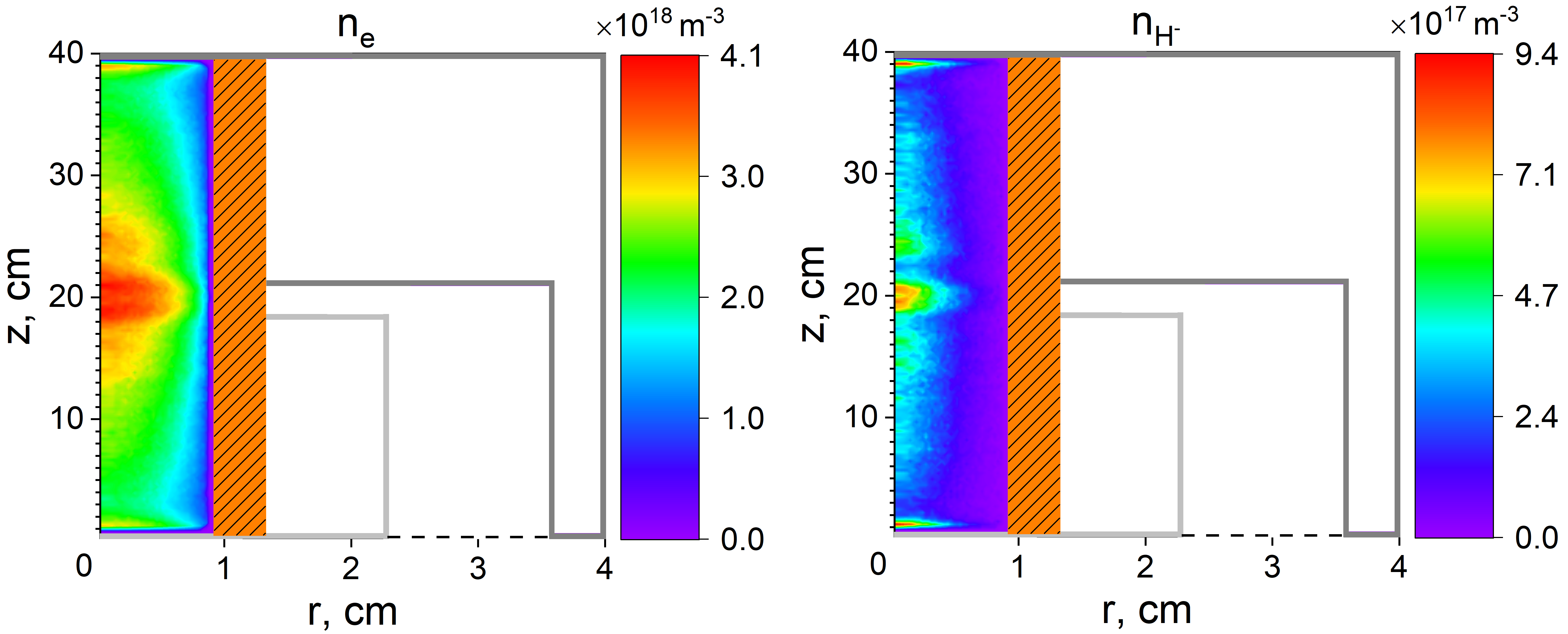}
\caption{Time-averaged electron and negative ion density obtained in PIC simulations of the surfatron discharge.}
\label{Fig3}
\end{figure}

\begin{figure}[h]
\centering
\includegraphics[width=12cm]{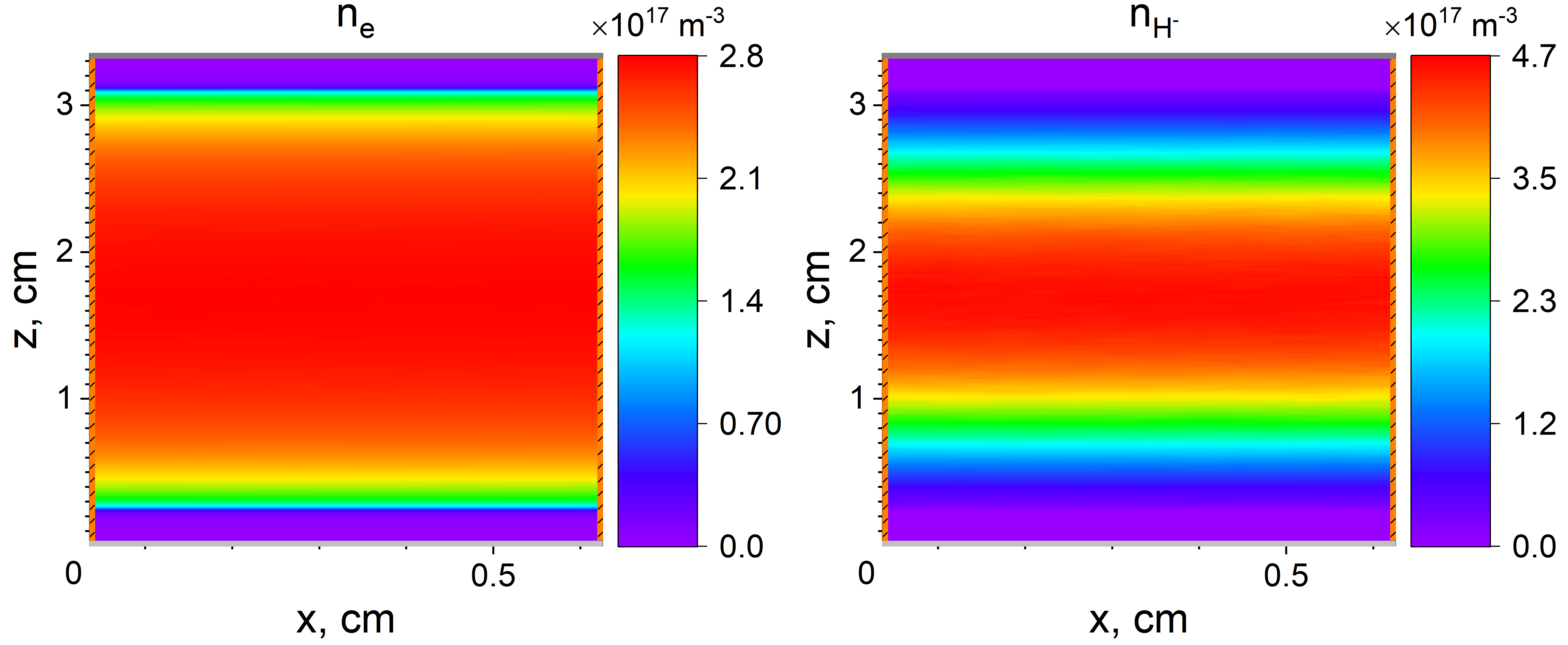}
\caption{Time-averaged electron and negative ion density obtained in PIC simulations of the CCP discharge.}
\label{Fig4}
\end{figure}

\begin{figure}[h]
\centering
\includegraphics[width=12cm]{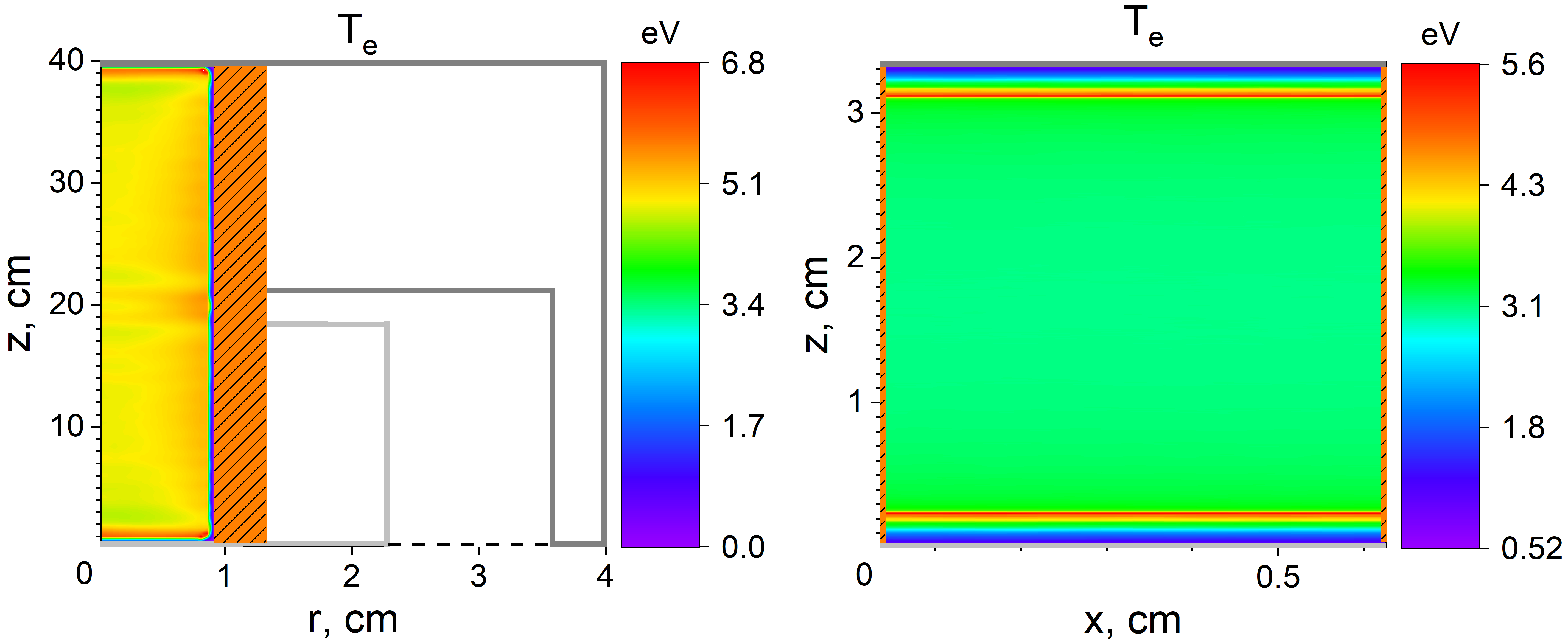}
\caption{Electron temperature obtained in PIC simulations of the surfatron (left) and the CCP (right) discharge.}
\label{Fig5}
\end{figure}

The aforementioned surfatron and CCP discharges are simulated for hydrogen at 300 K and 5 Pa. 

The power for the surfatron simulation is chosen so that the plasma would fill the entire reactor tube, which yields 4 kW. In Fig.\ref{Fig3}, left, one can see that the electron density is indeed quite high, reaching about $4 \times 10^{18}$ m$^{-3}$ at the location where the electromagnetic wave driven at 2.45 GHz enters the reactor chamber and gets converted into a surface wave propagating axially along the dielectric tube. Then, electron density falls off axially, which agrees with experimental observations. However, there is a local electron density elevation close to the end plates, which can be attributed to some edge effects (e.g., sheath heating). Fig.\ref{Fig3}, right, demonstrates that the density of $H^{-}$ ions is relatively low, with the maximum coinciding with the maximum of electron density, and the former being about 23\% of the latter. Note that the density profiles are strongly nonuniform both in axial and radial directions.

In contrast, CCP discharge driven at 200 MHz by a voltage with amplitude of 1.5 kV exhibits horizontally uniform density profiles for all species, see Fig.\ref{Fig4} for electron and negative ion density plots. Another difference from the surfatron discharge is that the negative ion density is nearly twice the electron density, except in the regions near the electrodes, where electrons are the dominant negatively charged species. Such a strongly ionic plasma in the CCP discharge can be attributed to a significantly colder average bulk electron energy in the CCP discharge ($\approx$ 3 eV) compared to that in the surfatron discharge ($\approx$ 5 eV), see Fig. \ref{Fig5}. This behavior can be explained by the characteristic features of low-pressure CCP discharges \cite{godyak_1990}, in which energetic electrons capable of causing ionization are efficiently generated near the moving sheath edge. These electrons then traverse the bulk plasma, occasionally undergoing ionizing collisions that produce additional electrons. However, the bulk electron population is confined by the ambipolar potential and can remain cold, because the aforementioned energetic electrons needed for the discharge sustainment come in sufficient numbers from the geometrically separated plasma sheath region. Such a regime with cold electron bulk is triggered when the electron mean path becomes comparable to the electrode distance. Although the electron temperature in the bulk of discharges operating in Ramsauer gases can become so low as the room temperature, because of the inefficient collisional heating of electrons there due to the low collisionality at low energies owing to the Ramsauer minimum, a reduced electron temperature can be observed in other gases, too. As argued before, lower energy of electrons leads to the weakening of the electron detachment mechanism destroying $H^{-}$, which results in the observed high density of negative ions. At the same time, the energetic tail of the electron energy distribution contains a sufficient population of electrons with energies above 20 eV to sustain vibrational excitation via the EV process, thereby facilitating substantial negative ion production. 

\vskip 0.5cm
\section{Conclusions \label{s4}}

In this work, implicit energy- and charge-conserving PIC/MCC simulations were used to investigate negative hydrogen ion production in two types of hydrogen discharges: a microwave-driven surfatron and an RF-driven capacitively coupled plasma (CCP), both operating at 5 Pa. The simulations incorporated a hydrogen chemistry model, which was enhanced from a previous study of the author to account for negative ion production and destruction. 

The results demonstrate substantial differences in negative ion production between the two discharge configurations. Most importantly, in the surfatron discharge, the electron density reaches approximately $4 \times 10^{18}$ m$^{-3}$, while the maximum H$^{-}$ density is about 23\% of the electron density. In contrast, the CCP discharge has nearly uniform density profiles in the horizontal direction, with the H$^{-}$ density in the bulk plasma exceeding the electron density by almost a factor of two. This difference is attributed primarily to the lower average bulk electron energy in the CCP discharge, approximately 3 eV compared with 5 eV in the surfatron. The resulting reduction in electron detachment suppresses H$^{-}$ losses and promotes negative ion accumulation.

These findings highlight the importance of electron energy distributions and discharge-specific heating mechanisms in determining negative ion densities in hydrogen plasmas. The applied energy-conserving PIC/MCC approach provides a self-consistent kinetic framework for investigating these processes under realistic discharge conditions. 




\vskip 0.5cm

\bibliographystyle{ieeetr}

\bibliography{references}




\end{document}